\documentclass[prc,twocolumn,superscriptaddress]{revtex4-1}
\usepackage{epsfig}
\usepackage{graphicx}
\usepackage{palatino}
\usepackage[english]{babel}
\usepackage{hyphenat}
\usepackage{amsmath}
\usepackage{amssymb}
\usepackage{slashed}
\usepackage{epstopdf}
\usepackage{xcolor}
\usepackage{booktabs}
\definecolor{lcolor}{rgb}{0.,0.0,0.}
\definecolor{citcolor}{rgb}{0,0.,0.5}
\usepackage[breaklinks,colorlinks,urlcolor=blue,citecolor=blue,linkcolor=blue]{hyperref}
\usepackage{multirow}
\usepackage{ltablex}

\newcommand{\beq}{\begin{equation}}
\newcommand{\eeq}{\end{equation}}
\newcommand{\bea}{\begin{eqnarray}}
\newcommand{\eea}{\end{eqnarray}}

\def\dd{{\rm d}}

\newcommand{\bem}{\begin{multline}}
\newcommand{\eem}{\end{multline}}
\newcommand{\beg}{\begin{gather}}
\newcommand{\eeg}{\end{gather}}

\newcommand{\nn}{\nonumber\\}

\newcommand{\ben}{\begin{eqnarray*}}
\newcommand{\een}{\end{eqnarray*}}

\newcommand{\bal}{\begin{align}}
\newcommand{\eal}{\begin{align}}

\newcommand{\eqn}[1]{Eq.~\eqref{#1}}
\newcommand{\fign}[1]{Fig.~\ref{#1}}
\newcommand{\secn}[1]{Section~1}
\newcommand{\appn}[1]{Appendix~1}

\long\def\comment#1{ }

\def\and{\quad\text{and}\quad}

\newcommand{\rmd}{{\rm d}}

\newcommand{\rme}{{\rm e}}

\def\0{{\boldsymbol 0}}

\def\ttaufull{\kappa}
\def\ttau{\kappa}
\def\max{{\rm max}}

\newcommand{\abar}{\bar{\alpha}}

\begin{document}

\title{Dynamical grooming of QCD jets}
\author{Yacine Mehtar-Tani}
\email[]{mehtartani@bnl.gov}
\affiliation{Physics Department, Brookhaven National Laboratory, Upton, NY 11973, USA}
\author{Alba Soto-Ontoso}
\email[]{ontoso@bnl.gov}
\affiliation{Physics Department, Brookhaven National Laboratory, Upton, NY 11973, USA}
\author{Konrad Tywoniuk}
\email[]{konrad.tywoniuk@uib.no}
\affiliation{Department of Physics and Technology, University of Bergen, 5007 Bergen, Norway}

\begin{abstract}
We propose a new class of infrared-collinear (IRC) and Sudakov safe observables with an associated jet grooming technique that removes dynamically soft and large angle branches. It is based on identifying the \textit{hardest} branch in the Cambridge/Aachen re-clustering sequence and discarding prior splittings that occur at larger angles. This leads to a dynamically generated cut-off on the phase space of the tagged splitting that is encoded in a Sudakov form factor. In this exploratory study we focus on the mass and momentum sharing distributions of the tagged splitting which we analyze analytically to modified leading logarithmic accuracy and compare to Monte-Carlo simulations. 
\end{abstract}

\maketitle

\section{Introduction}
\label{sec:intro}

Jets, or collimated sprays of particles originating from the fragmentation of energetic quarks and gluons, are among the most prominent features of high-energy particle collisions. The analysis of jet observables is crucial to study the theory of strong interactions, QCD, in the perturbative regime, including the running of the strong coupling constant $\alpha_s$. These phenomena also play an important role in constraining the background in searches for heavy particles, including the Higgs boson \cite{Butterworth:2008iy} and particles beyond the Standard Model \cite{Aad:2015owa}.

In the case of high-energy hadronic collisions, however, the observables are strongly affected by a wide range of processes that are hard to account for in perturbation theory and conventional resummation techniques. These include radiation from outside of the jet (non-global logarithms) \cite{Dasgupta:2001sh} and non-perturbative effects such as hadronization and underlying event activity.
In the last decade, tackling these challenges has lead to an improved analytical understanding of jet substructure, see \cite{Larkoski:2017jix,Asquith:2018igt,Marzani:2019hun} for recent reviews, coinciding with the maturing of fast and versatile jet reclustering procedures \cite{Ellis:1993tq,Dokshitzer:1997in,Cacciari:2008gp}. 

In this context, several jet grooming techniques designed to reduce the jet's sensitivity to non-local and non-perturbative physics have been developed. Such techniques have further evolved toward being easier to interpret in terms of perturbative QCD \cite{Dasgupta:2013ihk,Dasgupta:2013via}. Representative examples are the modified Mass Drop (mMDT) grooming \cite{Dasgupta:2013ihk} and SoftDrop (SD) grooming \cite{Larkoski:2014wba} that provide a two-parameter algorithm to determine the first  branching in an angular-ordered tree that is deemed to be sufficiently perturbative. 
Given the $i$-th primary emission off an angular ordered jet (corresponding to the $i$-th branch of a Cambridge/Aachen reclustering along the leading flow of energy), where $p_{{\rm T},i_1} \!>\! p_{{\rm T},i_2}$ are the energies of the two splitting products, one removes such emissions until one identifies the first whose momentum sharing fraction satisfies the condition $z\!>\!z_{\rm cut} (\theta/R)^\beta$.
Grooming recursively along the primary and secondary emission branches strongly reduces non-perturbative effects in specific cases \cite{Dreyer:2018tjj}. Other techniques, such as trimming \cite{Krohn:2009th}, recluster the jet with a smaller cone size and remove substructures below a certain energy cut-off.
Grooming techniques have also proven promising to study the internal structure of quenched jets in the context of heavy-ion collisions~\cite{Sirunyan:2017bsd,Kauder:2017mhg,Acharya:2018uvf,Acharya:2019djg}, see also \cite{Andrews:2018jcm}.

Furthermore, substructure techniques probe our knowledge of the multi-particle regime of QCD. Thus, designing new observables often goes hand in hand with a grooming scheme that permits a direct comparison to experimental data. For example, the momentum sharing variable $z_g$ of the first accepted emission in Soft Drop with $\beta\!=\!0$, that coincides with mMDT grooming \cite{Dasgupta:2013ihk}, turns out to be an ultraviolet fixed point \cite{Larkoski:2015lea} so that its distribution does not depend on the strong coupling constant. Other examples, such as the groomed jet mass \cite{Marzani:2017kqd,Kang:2018jwa}, have been calculated to next-to-leading logarithmic accuracy.

Despite the many successes, current jet substructure techniques are often quite simple but lacking an internal ``logic'' that would allow to estimate the most natural choice for the grooming parameters. 
These procedures are sensitive to the choice of parameters, e.g. $z_{\rm cut}$ and $ \beta$ in the case of SD, and their optimal values, in terms of resilience to underlying event or other distortions, can possibly depend on jet $p_{\rm \tiny T}$, underlying event activity and other unknown parameters. Clearly, if $z_{\rm cut} \ll 1$ the sensitivity to non-perturbative infrared effects is enhanced. Moreover, from an analytical point of view, their inclusion generates new scales on the level of jet substructure observables that complicate the understanding of the different contributing modes. This appears because the intrinsic jet scale is fluctuating on a jet-by-jet basis. One would therefore wish for a method that aligns more closely with the intrinsic properties of a given jet without the need for fine tuning.

\begin{figure}
\centering
\includegraphics[width=0.95\columnwidth]{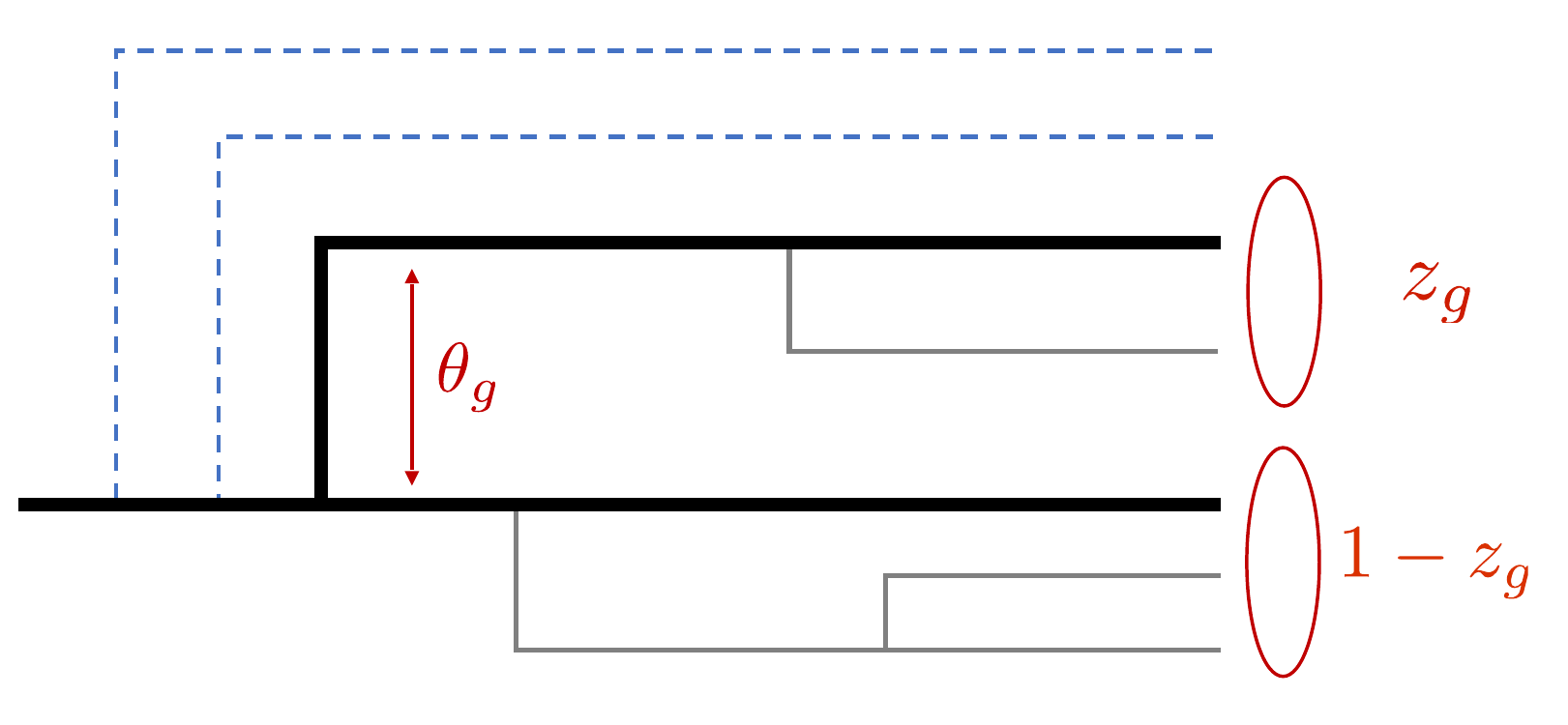}
\caption{Dynamical grooming applied to an angular ordered tree. The splitting represented by the thick (black) line has the largest $\ttau^{(a)}$ in the tree. In tagging mode, observables are calculated using the kinematic variables $z_g$ and $\theta_g$ of the tagged splitting. In  grooming mode, softer splittings that appear earlier in the tree, i.e. at larger angles, are discarded and the jet kinematics is adjusted accordingly.}
\label{fig:illustration}
\end{figure}
In this work, we aim to alleviate some of these shortcomings.
We consider a class of observables based on selecting, or tagging, the \textit{hardest} splitting in an angular ordered shower, where the ``hardness'' is characterized by a pseudo energy correlation variable as follows. Given the  $i$-th $1\!\to\!2$ splitting in the C/A~\cite{Dokshitzer:1997in} re-clustering sequence, the variable that measures the ``hardness'' of a jet, or in other words defines the ``hardest'' splitting within a jet, is defined as 
\begin{align}
\label{eq:hardness}
\ttaufull^{(a)} =\frac{1}{p_{\rm T}}\,\,\underset{i\in\, \text{C/A seq.}}{\max}\left[z_i(1-z_i) \,p_{{\rm T},i} \left( \frac{\theta_i}{R} \right)^a  \right]\,,
\end{align}
where $p_{\rm T}$ and $R$ are the energy and radius of the jet respectively, $z_i$ is the momentum sharing fraction, $p_{{\rm T},i}$ the energy of the parent, $\theta_i$ the relative angle of the splitting and $a$ is a free parameter whose physical interpretation will be discussed below.

Note that for $a\!=\!2$, we would select the splitting with the shortest formation time $t_{\rm f}^{-1} \!\sim\!\kappa^{(2)} p_{\rm T}$. We refer to this case as TimeDrop in what follows. Alternatively, for $a\!=\!1$ we tag the branching with the largest relative transverse momentum $k_{T}\!\sim\!\kappa^{(1)}\,p_{\rm T}$ and name this option as $k_{T}$Drop. Finally, $a\!=\!0$ corresponds to the splitting with the most symmetric momentum sharing and is called $z$Drop in what follows. In fact, $a\!=\!0$ leads to collinear sensitivity, see App.~\ref{appendix:a0-case}, and we will rather use $a\!=\!0.1$ for all practical purposes below.

Having identified a genuinely hard branching in the shower, we suggest two strategies. 
\begin{itemize}
\item In tagging mode, the kinematics of the hardest splitting informs the observable one wishes to compute. This will be the main focus of this paper.
\item In grooming mode, one discards all emissions taking place prior to the hard splitting in the re-clustering sequence. This procedure can easily be iterated along all the branches of the jet. We will pursue this strategy for other groomed observables in an upcoming publication.
\end{itemize} 

The main advantage of this method is that it autogenerates the conditions for tagging or grooming on a jet-by-jet basis. While a similar strategy is also pursued within e.g. jet pruning \cite{Ellis:2009su}, the procedure in our case is simpler to implement and closer in spirit to the physics of color coherence. In fact, softer emissions in the C/A sequence prior to the hardest one can be considered as radiation off the total charge. Our procedure only depends on one parameter which defines what we mean by the hardest emission inside a jet in contrast to most other techniques that involve two (extrinsic) parameters. Hence, we refer to this procedure as dynamical grooming.  A schematic illustration of how to dynamically groom an angular ordered shower can be found in Fig.~\ref{fig:illustration}.

This paper is structured as follows. In Sec.~\ref{sec:grooming}, we discuss vetoed showers and introduce the probability for a splitting to be the hardest. In Sec.~\ref{sec:observable}, we employ the derived Sudakov form factor to compute a family of observables based on the tagged splitting. We then perform analytical calculations in the modified leading-logarithmic approximation (MLLA) for the tagged mass and $z$ distributions in Sec.~\ref{sec:massdist} and Sec.~\ref{sec:zdist}. Finally, Sec.~\ref{sec:numerics} is dedicated to Monte-Carlo simulations of proton-proton collisions. We generate and analyze the Lund planes for the tagged splittings within the different dynamical grooming settings and present a systematic study of the impact of non-perturbative phenomena on the tagged mass and $z$ distributions. We end with a short discussion and outlook in Sec.~\ref{sec:conclusions}.

\section{Vetoed showers and tagging}
\label{sec:grooming}

Dynamically grooming a jet amounts to a certain reorganization of the conventional parton shower, where we will assume angular ordering. Given a specific $1\to 2$ splitting in the jet history, the procedure forces the emissions taking place both before, i.e. at angles larger than the selected splitting, and after, i.e. at smaller angles, not to allow for a harder emission. The reorganization depends on the properties of a given jet allowing for a procedure that is more adapted to account for jet-by-jet fluctuations.

The "hardness'' variable $\kappa^{(a)}$ is easily accessible in experimental data together with Monte-Carlo showers, and will be used in Sec.~\ref{sec:numerics}. In order to tackle the problem analytically, in a transparent manner and up to the required logarithmic precision, some simplifications are in order. To next-to-leading logarithmic accuracy in an angular ordered shower, it can be shown that the hardest splitting takes place off the leading particle in the jet or, in other words, on the primary Lund plane \cite{Andersson:1988gp,Dreyer:2018nbf}.  In this case, we neglect the energy depletion of the leading jet $p_{{\rm T},i}\!\simeq \!p_{\rm T}$ and hence explicit dependence on momenta cancel out in \eqn{eq:hardness}.  Accordingly, for a collinear safe definition, i.e. for $a\!>\!0$ in \eqn{eq:hardness}, the hardness of a tagged splitting is given by
\beq
\label{eq:hardness-1}
\ttau^{(a)} = z(1-z) \left(\frac{\theta}{R} \right)^a \,.
\eeq
Whenever it is obvious from the context, we will simply write $\ttau \equiv \ttau^{(a)}$.

The central quantity in our computations is the following Sudakov form factor,
\begin{align}
\label{eq:sudakov-0}
\ln \Delta \big(\ttau \big|a) &= -  \int_0^{R} \frac{\dd \theta}{\theta} \int_0^1 \dd z\,\frac{\alpha_s(k^2_t)}{\pi} P(z) \nn
& \times \Theta \big( z(1-z) (\theta/R)^a > \ttau \big)\,,
\end{align}
where $P(z)$ is the splitting function and the Heaviside function vetoes emissions with a $\ttau^{(a)}$ larger than the measured, or tagged, emission. Note that the angular integral, spanning between 0 and $R$, enforces this veto all over the primary Lund plane of the jet. Finally, $\alpha_s(k^2_t)$ is the strong coupling constant evaluated at $k_t\!=\!z(1-z)\theta p_{\rm T}$, the transverse momentum generated at the splitting.
 
Such a form factor arises as a remainder contribution from the vetoed showers occurring before and after the hard emission. A similar construction was previously used as a method to match parton showers with next-to-leading order contributions where the hardest emission was the one with the largest $k_t \!\sim\! \kappa^{(1)}$ \cite{Nason:2004rx}. Here, we will proceed with a more direct line of reasoning to derive the relevant probability distribution.
Taking the derivative and then integrating over $\ttau$ in \eqn{eq:sudakov-0} leads to the following identity,
\beq
\int_0^\infty \rmd \ttau \, \frac{\rmd }{\rmd \ttau}\, \Delta(\ttau) = \Delta(\infty) - \Delta(0) \,
\eeq
where $\Delta(\kappa) \equiv \Delta(\kappa \big| a)$. 
Clearly, we have that $\Delta(\infty)\!=\!1$. For $a\!>\!0$, the Sudakov given in \eqn{eq:sudakov-0} is infrared and collinear finite. Therefore, we can safely take the limit $\ttau\!\to\!0$, resulting in $\Delta(0)\!=\!0$. 
This is no longer the case for collinear unsafe observables, explicitly for $a\!=\!0$, where we have to introduce a non-perturbative cut-off scale to regulate the integrals. We will treat this particular case in more detail in App.~\ref{appendix:a0-case}, and focus in the remainder of the paper on the collinear safe taggers.

Then, for collinear safe observables we can construct a normalized probability distribution of the splitting with the largest $\ttau^{(a)}$ in an angular ordered shower. 
Owing to the fact that 
\begin{align}
\label{eq:sudakov-eq}
\frac{\rmd }{\rmd \ttau} \Delta(\ttau) &= \int_0^{R} \frac{\rmd \theta}{\theta}\int_0^1\rmd z \, \frac{\alpha_s(k^2_t)}{\pi}P(z) \,\Delta(\ttau^{(a)} \big| a) \nn
&\times \delta (z(1-z) (\theta/R)^a-\ttau) \,,
\end{align}
it follows that 
\beq
\int_0^{R} \frac{\dd \theta}{ \theta} \int_0^1 \frac{\rmd z}{z}\, \mathcal{P}(z,\theta) =1,
\eeq
where 
\beq
\label{eq:prob-dist}
{\cal P}(z,\theta) = \frac{\alpha_s(k^2_t)}{\pi} \, zP(z)\, \Delta(\kappa \big| a)  \,,
\eeq
is the probability of splitting giving rise to the momentum sharing fraction $z$ at angle $\theta$ that results into the largest $\kappa^{(a)}$ in the shower, i.e. to be the hardest splitting.

In order to gain analytic insight, we will work for simplicity in the modified leading logarithmic approximation that assumes angular ordering and where the one loop Altarelli-Parisi quark-gluon splitting function can be approximated as
\beq
\label{eq:split-1}
P_{gq}(z) = C_F\frac{1+(1-z)^2}{z} \approx \frac{2C_F}{z} \left(1 - \frac{3}{4}z \right) \,,
\eeq
where the second term comes from approximating the finite part of the splitting function by $\int_0^1 \rmd z\,[P(z)/C_F-2/z]\!=\!-3/2$. This corresponds to the modified leading-logarithmic approximation.
For fixed coupling and using the approximation in \eqn{eq:split-1}, the integrals in \eqn{eq:sudakov-0} can be done analytically. The final expression for the Sudakov form factor in the MLLA is then,
\beq
\label{eq:sudakov-mlla-1}
\ln \Delta(\ttau\big|a) = - \frac{\bar \alpha}{a}\left[\ln^2 \ttau+ \frac{3}{2} \left(\ln \ttau + 1-\ttau\right) \right] \,,
\eeq
where $\bar \alpha \equiv \alpha_s C_F/\pi$. Only the first two terms are relevant to leading-logarithmic (LL) accuracy. For the numerical evaluations, we will keep the term of order one to insure exact normalization of our observables. An analogous derivation of the Sudakov form factor for gluon-initiated jets can be found in App.~\ref{appendix:gluon-case}.

\section{Computing tagged observables}
\label{sec:observable}
Since our procedure exploits the properties of the hardest branching in the jet shower, we will be interested in observables that are of the same form as \eqn{eq:hardness-1}. 
Generically, such distributions are then given by
\begin{align}
\label{eq:hardness-general}
\left.\frac{1}{\sigma} \frac{\dd\sigma}{\dd \ttau^{(b)}} \right|_a &= \int_0^{R} \frac{\dd\theta}{\theta} \int_0^1 \dd z \, \frac{\alpha_s(k^2_t)}{\pi} P(z) \Delta\big(\ttau^{(a)}\big| a\big)\nn
&\times \delta\big(z(1-z)(\theta/R)^b - \ttau^{(b)}\big) \,,
\end{align}
where $\sigma|_a$ represents the fact that the cross section distribution is measured given a tagged splitting with the largest $\ttau^{(a)}$ in the parton shower. The normalized distribution of the jet "hardness'' is then written as 
\beq
{\cal H}(b|a) \equiv \left.\frac{1}{\sigma} \frac{\dd\sigma}{\dd \ttau^{(b)}} \right|_a\,.
\eeq
Hence, the first argument of the function is related to the observable $\ttau^{(b)}$ that is measured on the kinematics of the tagged splitting while the second defines what do we mean by tagging the hardest splitting, i.e. we identify the splitting with the largest $\ttau^{(a)}$. 
Whenever it is obvious from the context, we will simply denote $\ttau^{(b)} \equiv \ttau$.

\begin{figure*}
\centering
\includegraphics[width=0.95\textwidth]{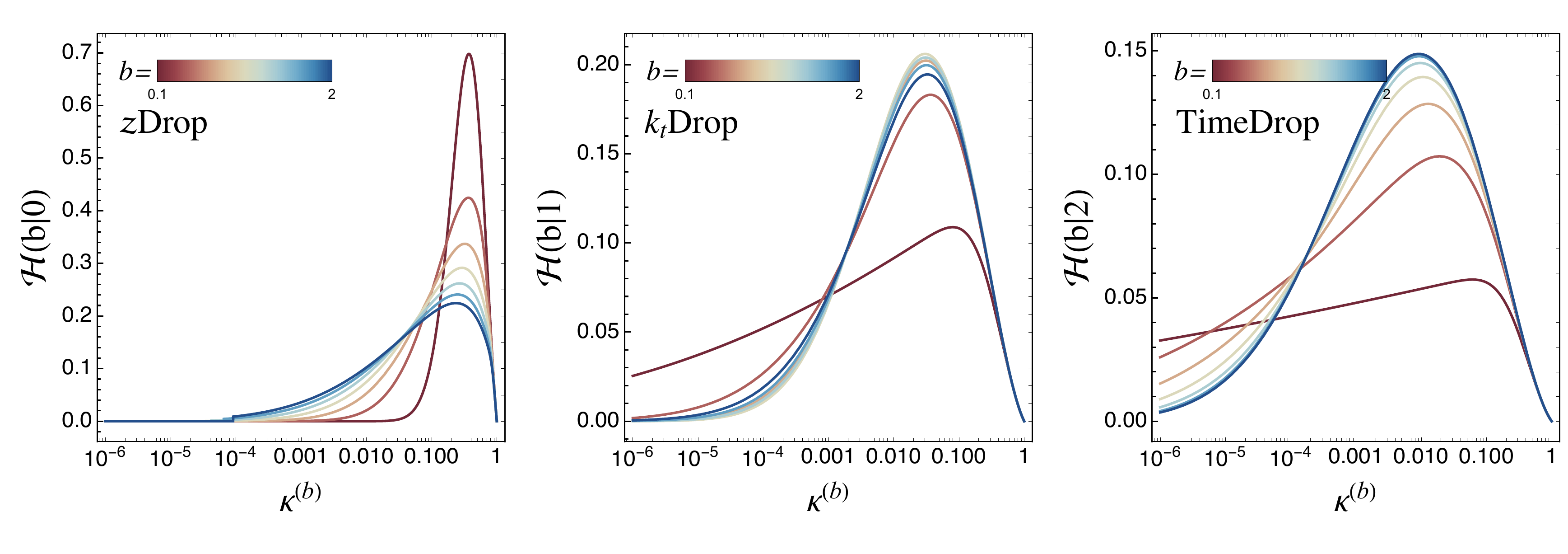}
\caption{The normalized distribution ${\cal H}(b \big| a)$ for three grooming modes: $z$Drop ($a=0.1$, left panel), $k_t$Drop ($a=1$, central panel) and TimeDrop ($a=2$, right panel). We have generated distributions for a range of $b$ values in Eq.~\ref{eq:hardness-mlla-0} for fixed coupling at MLLA.}
\label{fig:Hba-panel}
\end{figure*}

Without losing generality, we work in the MLL approximation and treat $z \!\ll\!1$. Then, assuming $b\!>\!0$, we evaluate the $\delta$-function to obtain
\begin{align}
\label{eq:hardness-mlla-0}
{\cal H}(b|a)&= \frac{1}{b}\int_{\ttau}^1\dd z \,  \frac{\alpha_s(k^2_t)}{\pi}\,P(z)\, \Delta\big(z^{1-\frac{a}{b}} \ttau^{\frac{a}{b}}|a\big) \,,
\end{align}
where now the argument of the running coupling is $k_t\!=\!z^{1-\frac{1}{b}} \ttau^{\frac{1}{b}} Q$ with  
$Q\equiv p_{\rm T} R$ being the jet scale. It is straightforward to check that \eqn{eq:hardness-mlla-0} is normalized to unity. 
After further simplifications, assuming $a \!\neq\!b$, the final expression reads
\beq
\label{eq:hardness-mlla}
{\cal H}(b|a) = \frac{1}{b-a}\int_{\ttau}^{\ttau^{\frac{a}{b}}}\frac{\dd x}{x} \, \frac{\alpha_s(k^2_t)}{\pi}\, \tilde P\left( x^{\frac{b}{b-a}} \ttau^{-\frac{a}{b-a}}\right) \, \Delta (x|a) \,,
\eeq
where we introduced the notation $\tilde P(z)\!\equiv\!z P(z)$ and where now $k_t\!=\!x^\frac{b-1}{b-a} \ttau^\frac{1-a}{b-a} Q$. 

It is clear from \eqn{eq:hardness-mlla} that, except when $b\!=\!0$, ${\cal H}(b|a) $ is an infrared and collinear safe quantity since it admits a Taylor expansion in the coupling constant.  For $b\!=\!0$, the $x$ integration goes from $0$ to $\kappa$ and the integrant exhibits an essential singularity at $0$ that is regulated by the Sudakov form factor. The resulting integral is an asymptotic series, and hence finite, although the perturbative expansion is divergent term by term.  In this case, ${\cal H}(0|a) $ is said to be Sudakov safe \cite{Larkoski:2013paa,Larkoski:2015lea}.

Remarkably, in the double logarithmic approximation (DLA), where we drop the second term in \eqn{eq:split-1}, and with fixed coupling, we obtain
\begin{align}
&{\cal H}(b|a) \simeq \frac{2 \bar \alpha}{b-a}\int_{\ttau}^{\ttau^{\frac{a}{b}}}\frac{\dd x}{x} \, \rme^{- \frac{\bar \alpha}{a} \ln^2 x} \nn
&\simeq \frac{\sqrt{\pi \, \bar\alpha a}}{b-a} \left[{\rm erf}\left(\sqrt{\frac{\abar a}{b^2}}\ln \ttau \right) -{\rm erf}\left(\sqrt{\frac{\abar}{a}}\ln \ttau \right) \right] \,,
\end{align}
which turns out to be invariant under the transformation $a \to b^2/a$, i.e.
\beq
\label{eq:duality}
{\cal H} \left(b \big|a \right) \approx {\cal H} \left(b \bigg| \frac{b^2}{a} \right)  \,,
\eeq
where the equality only holds at DLA. This immediately singles out $a\!=\!b$ as a ``fixed point'' of this class of observables. The dual distributions correspond to different tagging modes and in the limiting case $a\to 0$ this amounts to strong and weak grooming in the left and right hand side of \eqn{eq:duality}, respectively.  Below we will come back to the meaning of these observations.

We plot the normalized distribution ${\cal H}(b \big| a)$ as function of the variable $\ttau^{(b)}$ in Fig.~\ref{fig:Hba-panel} for three values of $a$, corresponding (from left to right) to $z$Drop ($a\!\approx\!0$), $k_t$Drop ($a\!=\!1$) and TimeDrop ($a\!=\!2$). For each grooming setting we plot the distributions for different variables $\ttau^{(b)}$ for $0.1\!<\!b\!<\!2$. To observe the approximate duality \eqn{eq:duality} it is sufficient to notice that, in the central panel ($k_t$Drop) the distributions reach the maximal values at $b\!=\!a$ and starts decreasing at $b\!>\!a $. For completeness, similar results for gluon-initiated jets are displayed in Appendix~\ref{appendix:gluon-case}.

In order to discuss the qualitative features of the spectrum at this level it is sufficient to focus on the case $b\!>\!a \!>\!0$. The asymptotic behavior of the  $\kappa$-distribution, i.e. $\ln^2 \ttau \gg  (\abar \, a)^{-1}$, reads 
\beq
\label{eq:hardness-dla-limits}
{\cal H}(b|a)  \approx \frac{b}{b-a}\frac{1}{\ln \ttau^{-1}} \rme^{-  \frac{\abar a}{b^2} \ln^2 \ttau}\,.
\eeq
We observe two qualitative features. First, the distribution peaks at $\ln \ttau \sim (\abar\,a/b^2)^{-1/2}$. Hence, the peak of the distribution shifts to larger values of $\kappa$ with decreasing $a$. Second, from \eqn{eq:hardness-dla-limits}, we see that the distribution flattens as $a$ decreases. The opposite case, i.e. $a\!>\!b\!>\!0$, can be found in a similar way, e.g. using \eqn{eq:duality} at DLA. These features are seen in Fig.~\ref{fig:Hba-panel}. As a result, the limits $a\to \infty$ and $a\to 0$ exhibit similar behavior despite their different groomed modes. 

Turning now to the special case, when $a\!=\! b \!>\!0$, we evaluate \eqn{eq:hardness-mlla-0} which gives
\beq
\label{eq:hardness-special}
{\cal H}(a|a) = \frac{1}{a} \Delta(\ttau|a) \int_{\ttau}^1 \dd z \,  \frac{\alpha_s(k^2_t)}{\pi} P(z) \,,
\eeq
where now $k_t\!=\!z^{1-\frac{1}{a}} \ttau^{\frac{1}{a}} Q$. The remaining integral in \eqn{eq:hardness-special} is regulated from below and is finite. In the MLLA and at fixed coupling we get
\beq
\label{eq:hardness-special-mll}
{\cal H}(a|a) = -\frac{2 \abar}{a}\left[  \ln \ttau  +\frac{3}{4}(1-\ttau)\right] \, \Delta(\ttau|a) \,.
\eeq
It turns out that the distribution measured this way corresponds to the plain distribution to LL accuracy, i.e. the distribution of the observables without any grooming, see \eqn{eq:MLLA-mass-2} for a concrete example. Hence, having $b\!\sim\!a$ corresponds to a low degree of grooming, such that the distribution closely resembles the plain one, while $b\!\gg\!a$ results in strong grooming.

We now proceed to consider in more detail two important observables in jet physics, namely the mass and momentum sharing fraction $z$ properties of the tagged, hard splitting. As a further example, we will consider the tagged $k_t$ distribution in App.~\ref{appendix:ktdist}.

\subsection{Tagged mass distribution ($b\!=\!2$)}
\label{sec:massdist}
 The case $b\!=\!2$ is related to the mass of a given splitting or, in other words, to the virtuality $m^2$ of the parent particle that decayed. Defining the rescaled variable $\rho \equiv m^2/(p_{\rm T} R)^2$, the normalized distribution is simply ${\cal H}(2 \big| a) \equiv \rmd \ln \sigma \big/ \rmd \ln \rho$. Using \eqn{eq:hardness-mlla-0}, we find at MLL accuracy,
\begin{align}
\label{eq:MLLA-mass}
{\cal H}(2\big|a) & = \frac{2\bar \alpha}{2-a} \int_{\ttau}^{\ttau^{a/2}} \frac{\dd x}{x}\,\left[1- \frac{3}{4} \frac{x^{\frac{2}{2-a}}}{\ttau^{\frac{a}{2-a}}} \right] \nn
&\times \exp\left\{-\frac{\bar \alpha}{a} \left[\ln^2 x+ \frac{3}{2}(\ln x+1-x) \right] \right\} \,,
\end{align}
where in this case $\ttau\!\equiv \!\ttau^{(2)} \!=\! \rho\!=\! z (1-z) (\theta/R)^2$. 

\begin{figure}
\includegraphics[width=\columnwidth]{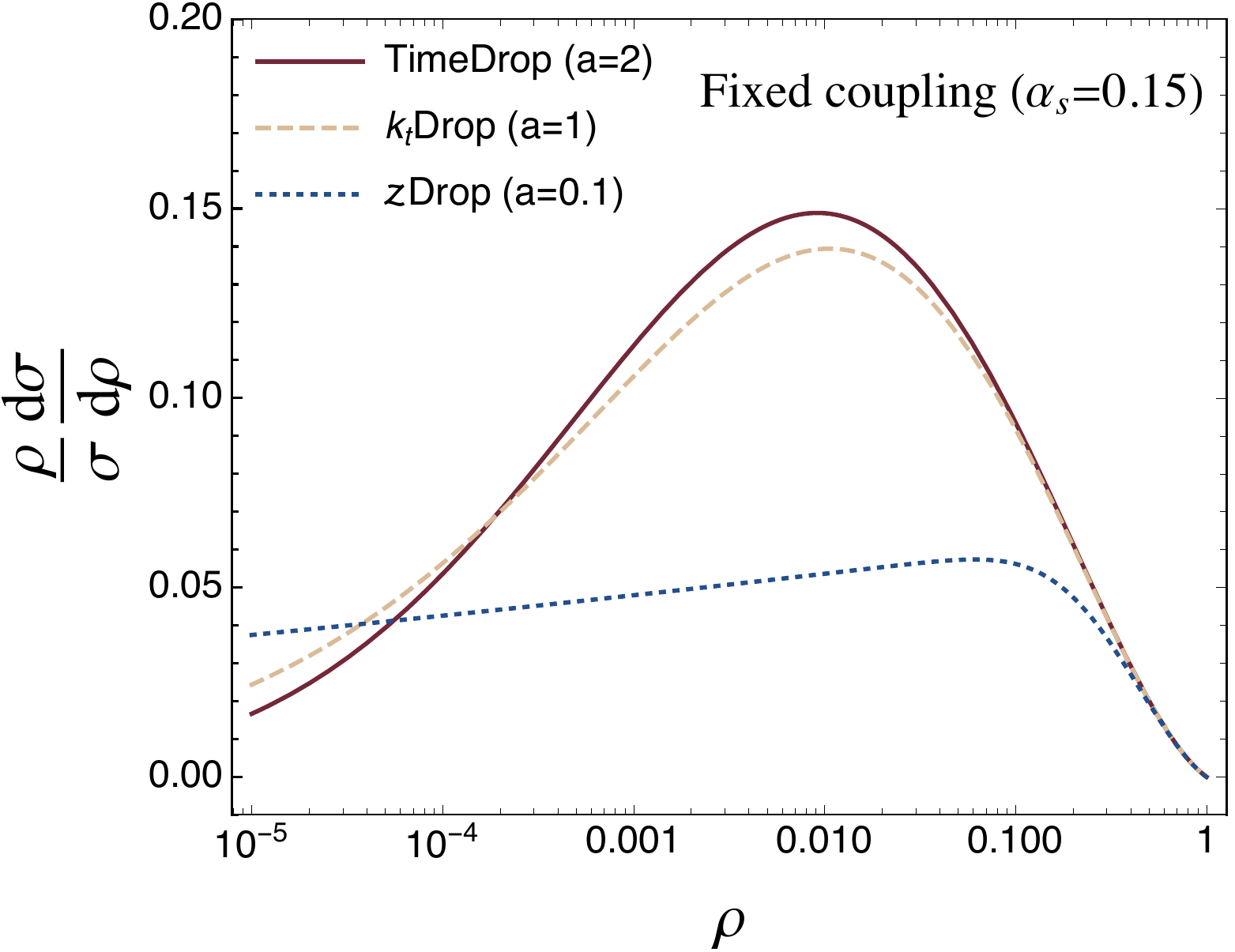}
\caption{The tagged mass distribution for fixed coupling as given by Eq.~\ref{eq:MLLA-mass} for $2\!>\!a\!>\!0$.}
\label{fig:mass-fc}
\end{figure}

The result of numerically solving \eqn{eq:MLLA-mass} is displayed in Fig.~\ref{fig:mass-fc} for the most representative values of $a$, i.e. TimeDrop ($a\!=\!2$), $k_t$Drop ($a\!=\!1$) and $z$Drop ($a\!\approx\!0$). We observe how the qualitative features generically discussed in the previous section are manifest: as $a$ decreases the $\rho$ distribution flattens while the peak shifts to larger values of $\rho$. 

The case where $b\!=\!a\!=\!2$ is of special interest. Using \eqn{eq:hardness-special-mll} we obtain
\begin{align}
\label{eq:MLLA-mass-2}
\frac{\rho}{\sigma} \frac{\dd \sigma}{\dd \rho} &=-\abar\, \left( \ln \rho+\frac{3}{4}(1-\rho) \right)\,\nn
&\times \exp\left\{-\bar \alpha \left[\frac{1}{2}\ln^2 \rho+ \frac{3}{4}(\ln \rho+1-\rho) \right] \right\} \,,
\end{align}
which remarkably reproduces the result for the plain mass at leading logarithmic accuracy.

This is not surprising since to this level of accuracy the plain mass is determined by the hardest splitting. 

\subsection{Tagged $z$ distribution ($b\!=\!0$)}
\label{sec:zdist}
\begin{figure}

\includegraphics[width=\columnwidth]{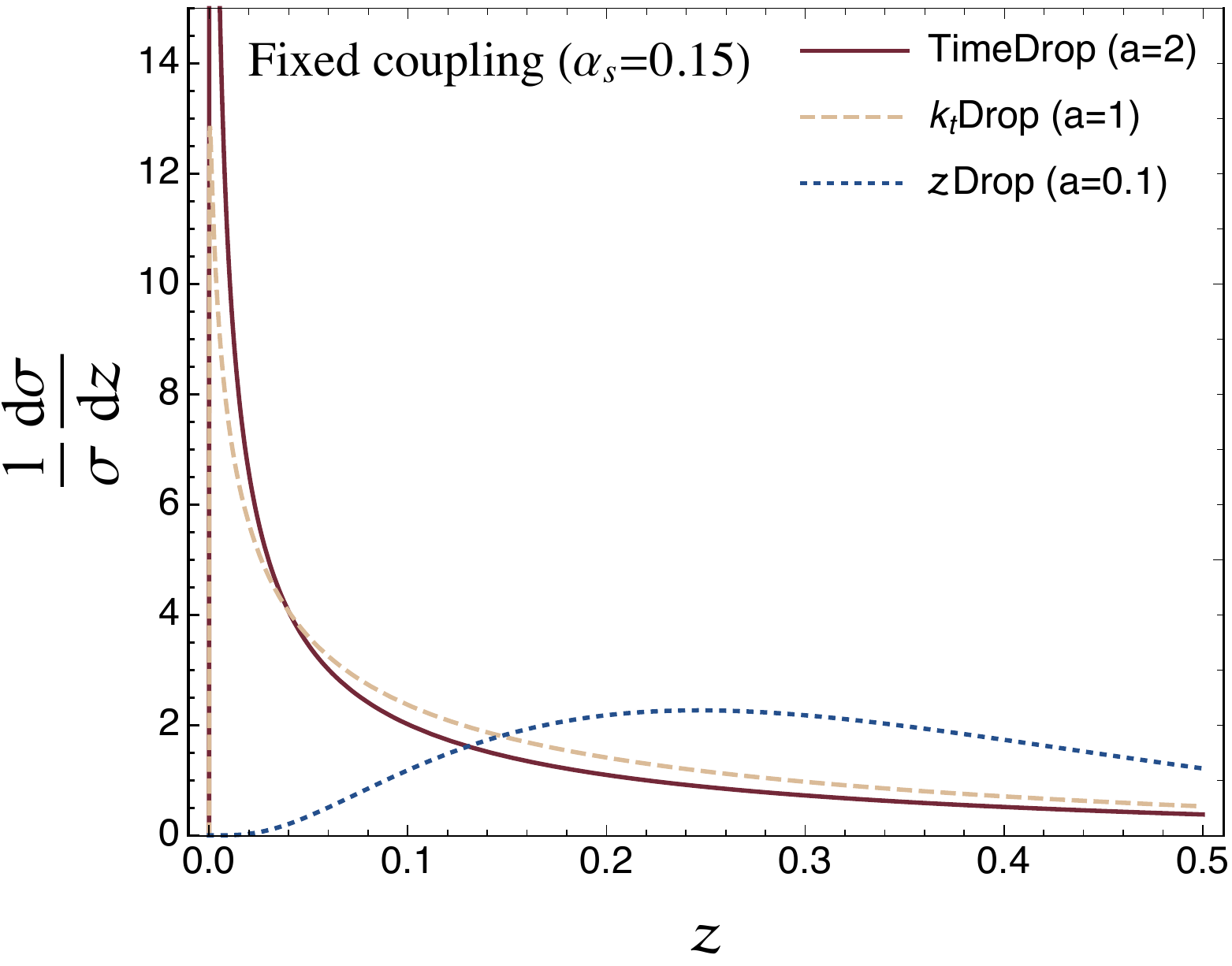}
\caption{The tagged $z$-distribution for fixed coupling as given by Eq.~\ref{eq:MLLA-z} for $2\!>\!a\!>\!0$.}
\label{fig:zdist-1}
\end{figure}
As a second example, we consider the tagged $z$-distribution, with $b\!=\!0$ so that $\ttau^{(0)}\!=\!z(1-z)\!\approx\!z$. Since we are now dealing with a potentially infrared unsafe but Sudakov safe~\cite{Larkoski:2015lea} observable, see \eqn{eq:hardness-general}, one has to beware. However, defining ${\cal H}(0 \big| a) \equiv \rmd \ln \sigma \big/ \rmd \ln z$, it straightforward to derive
\beq
\label{eq:gen-z}
\displaystyle\frac{1}{\sigma}\displaystyle\frac{\dd\sigma}{\dd z} 
= P(z)\displaystyle\int_0^{R} \displaystyle\frac{\dd\theta}{\theta}\,\frac{\alpha_s(k^2_t)}{\pi} \Delta(\ttau\big|a).
\eeq

\begin{figure*}
\includegraphics[width=\textwidth]{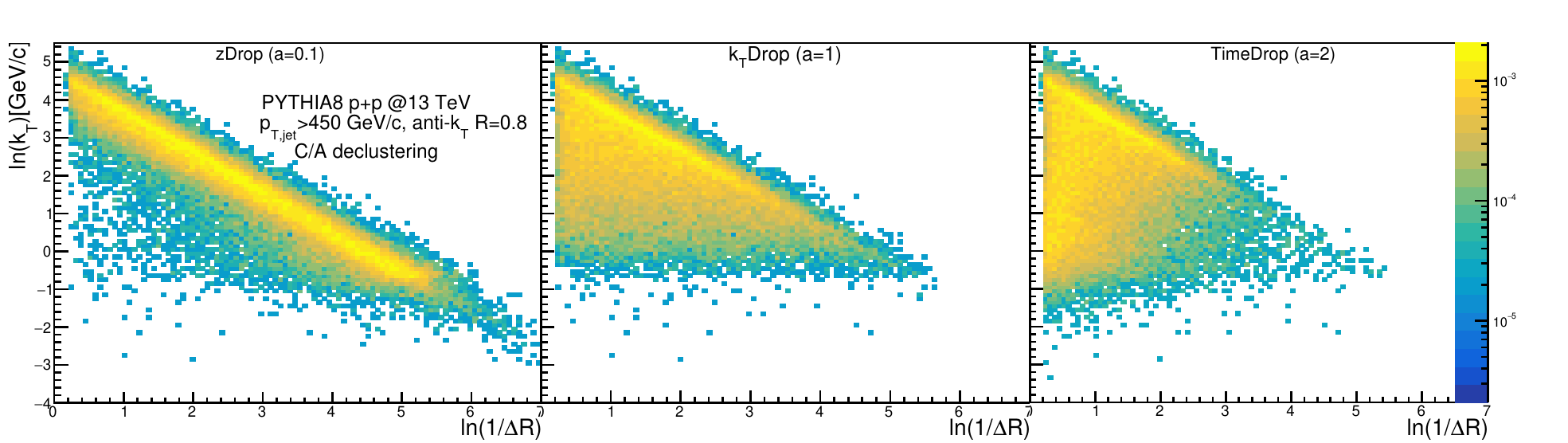}
\caption{Primary Lund planes for the tagged emissions. }
\label{fig:lundplane-dyg}
\end{figure*}

After fixing the coupling and at MLLA, \eqn{eq:gen-z} transforms into 
\begin{align}
\label{eq:MLLA-z}
\displaystyle\frac{1}{\sigma}\displaystyle\frac{\dd\sigma}{\dd z} &=  \frac{2\bar\alpha}{z\, a}\left ( 1-\displaystyle\frac{3}{4}z\right)\displaystyle\int_0^{z} \displaystyle\frac{\dd x}{x}\, \rme^{-\frac{\bar \alpha}{a} \left[\ln^2 x+ \frac{3}{2}(\ln x+1-x) \right]}  .
\end{align}
The resulting tagged $z$-distributions obtained by numerically solving \eqn{eq:MLLA-z} are displayed in Fig.~\ref{fig:zdist-1} for $2\!>\!a\!>\!0$. 
The origin of the main features observed in Fig.~\ref{fig:zdist-1} can be understood analytically by resorting to the DLA, where
\begin{align}
\frac{1}{\sigma} \frac{\dd \sigma}{\dd z}  &\approx \frac{2 \bar \alpha}{z \,a} \int_0^z \frac{\dd x}{x} \rme^{-\frac{\bar \alpha}{a}\ln^2 x} \nn
&\approx \frac{1}{z}\sqrt{\frac{\bar \alpha\pi}{a}} \left[{\rm erf} \left( \sqrt{\frac{\bar \alpha}{a}} \ln z\right)-1 \right] \,.
\end{align}
This distribution is cut-off at a characteristic value of $z$, namely
\beq
\label{eq:z-cutoff}
z_{\rm cut} \approx \rme^{-\sqrt{a/\bar \alpha}} \,.
\eeq
For $a \gg \abar $, this opens a wide range $z_{\rm cut}\!<\!z \!< 0.5$ where the distribution falls off as $z^{-1}$ and is modulated by $\sqrt{\bar \alpha/a}$. However, for $a \gtrsim 0$ and $z_{\rm cut} \approx 1$, we find that 
\beq
\frac{1}{\sigma} \frac{\dd \sigma}{\dd z} \sim \frac{1}{z} \rme^{- \frac{\bar \alpha}{a} \ln^2 z} \,,
\eeq
i.e. the distribution grows slowly with $z$. These features are roughly reproduced in \fign{fig:zdist-1} where the drop-off for the $k_t$Drop case is clearly visible around $z\!\sim\!0.02$.

In this context it is interesting to notice that the cut-off in $z$, \eqn{eq:z-cutoff}, is dynamically generated and is a measure of $\sqrt{\alpha_s/a}$. This is quite different from SD (mMDT) grooming with $\beta\!=\!0$ where the cut-off is simply given by the input to the algorithm. Although the distribution is modulated by the same ratio, dynamical grooming opens up the possibility to probe the splitting function down to low $z$.

\section{Monte-Carlo studies and resilience to non-perturbative effects}
\label{sec:numerics}
%

\begin{figure}
\centering
\includegraphics[width=0.95\columnwidth]{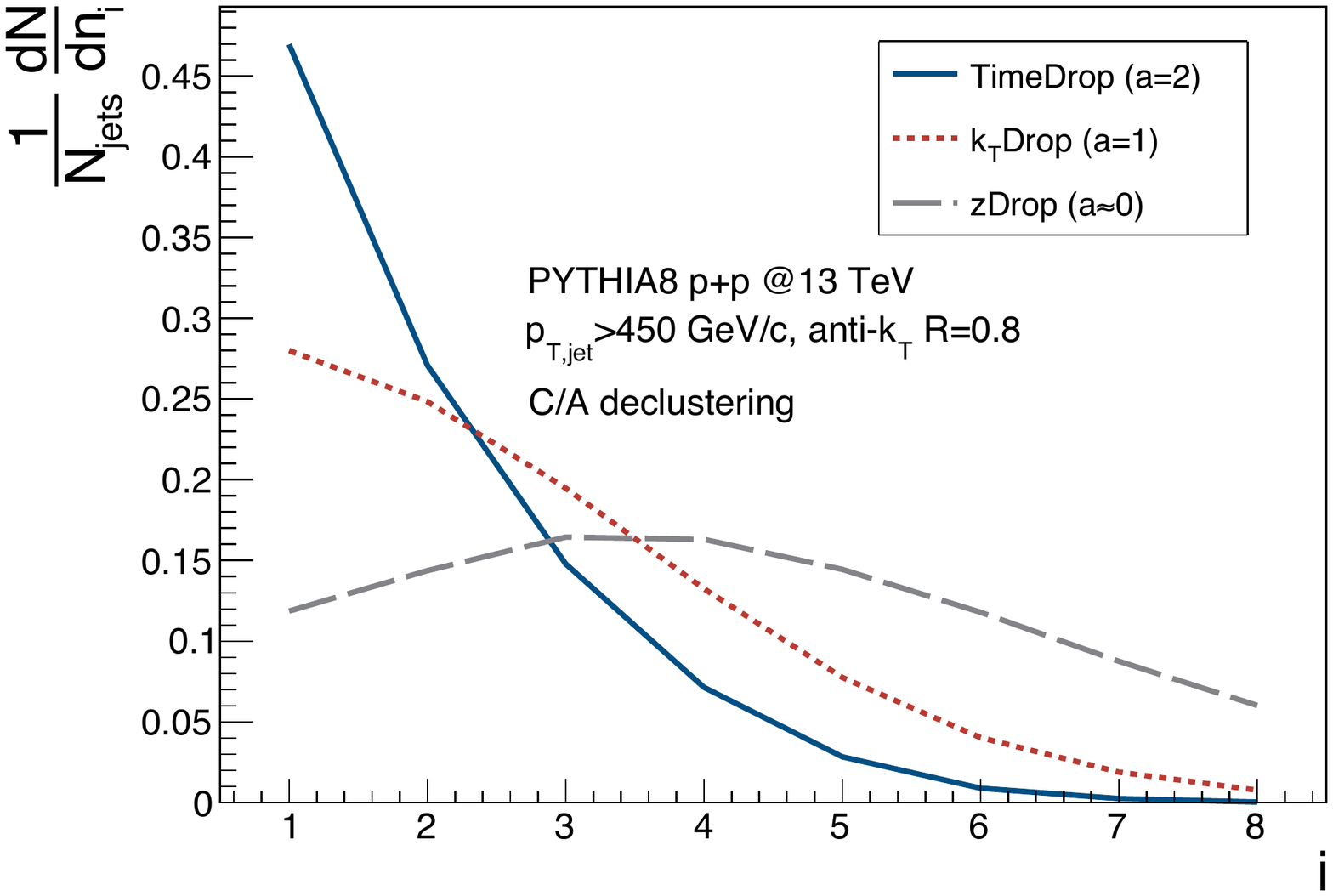}
\caption{Distribution of the number of the ``hardest'' branching on the primary Lund plane.}
\label{fig:n-hardest}
\end{figure}

\begin{figure}[h]
\includegraphics[width=\columnwidth]{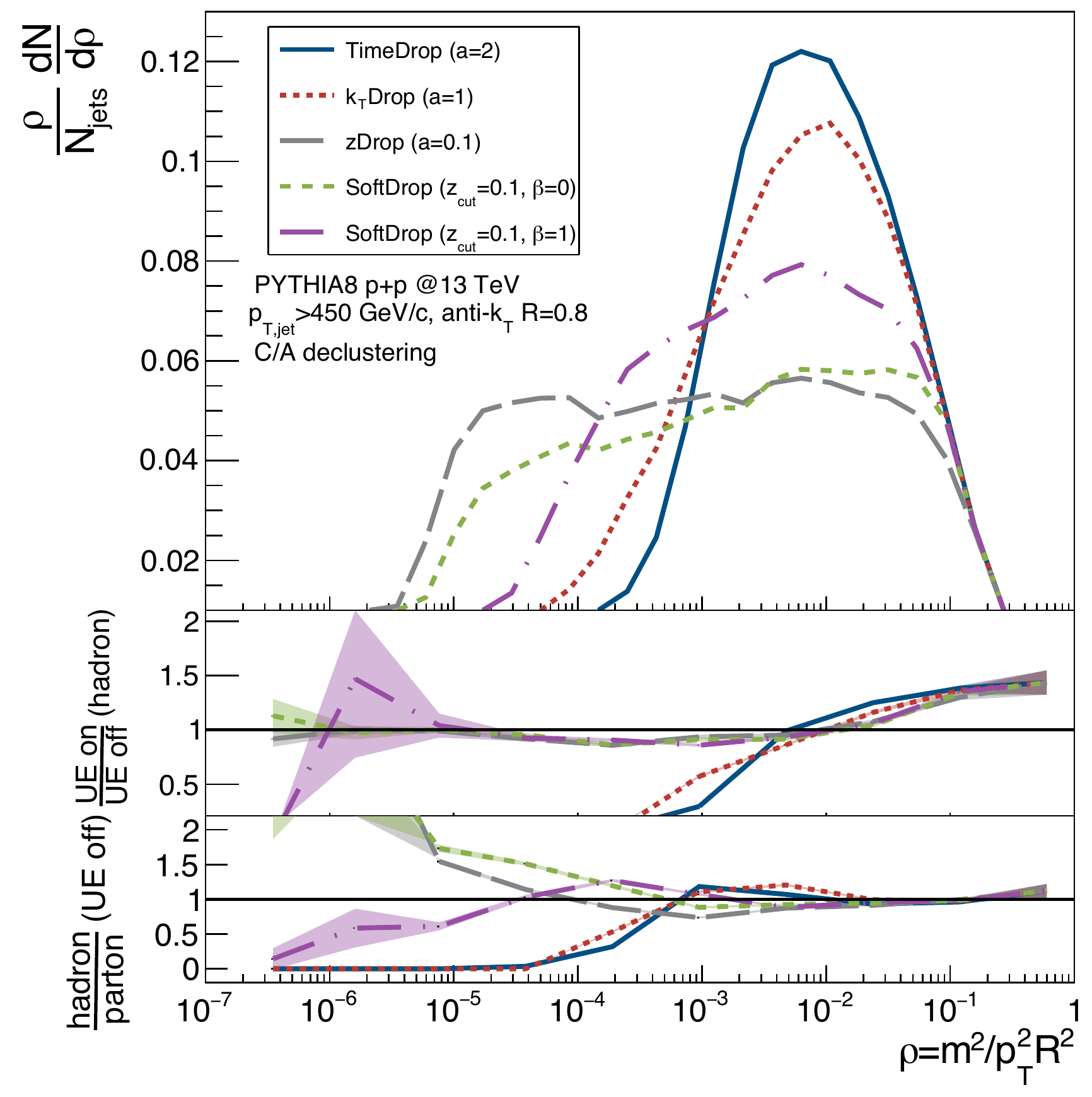}
\caption{Top: the tagged $\rho$-distribution for different choices of $a$ in Eq.~\ref{eq:hardness} and SoftDrop. Middle: ratio of the distributions with and without underlying event at hadron level. Bottom: ratio of the distributions at hadronic and partonic level without underlying event.}
\label{fig:rhodist-mc}
\end{figure}

\begin{figure}[h]
\includegraphics[width=\columnwidth]{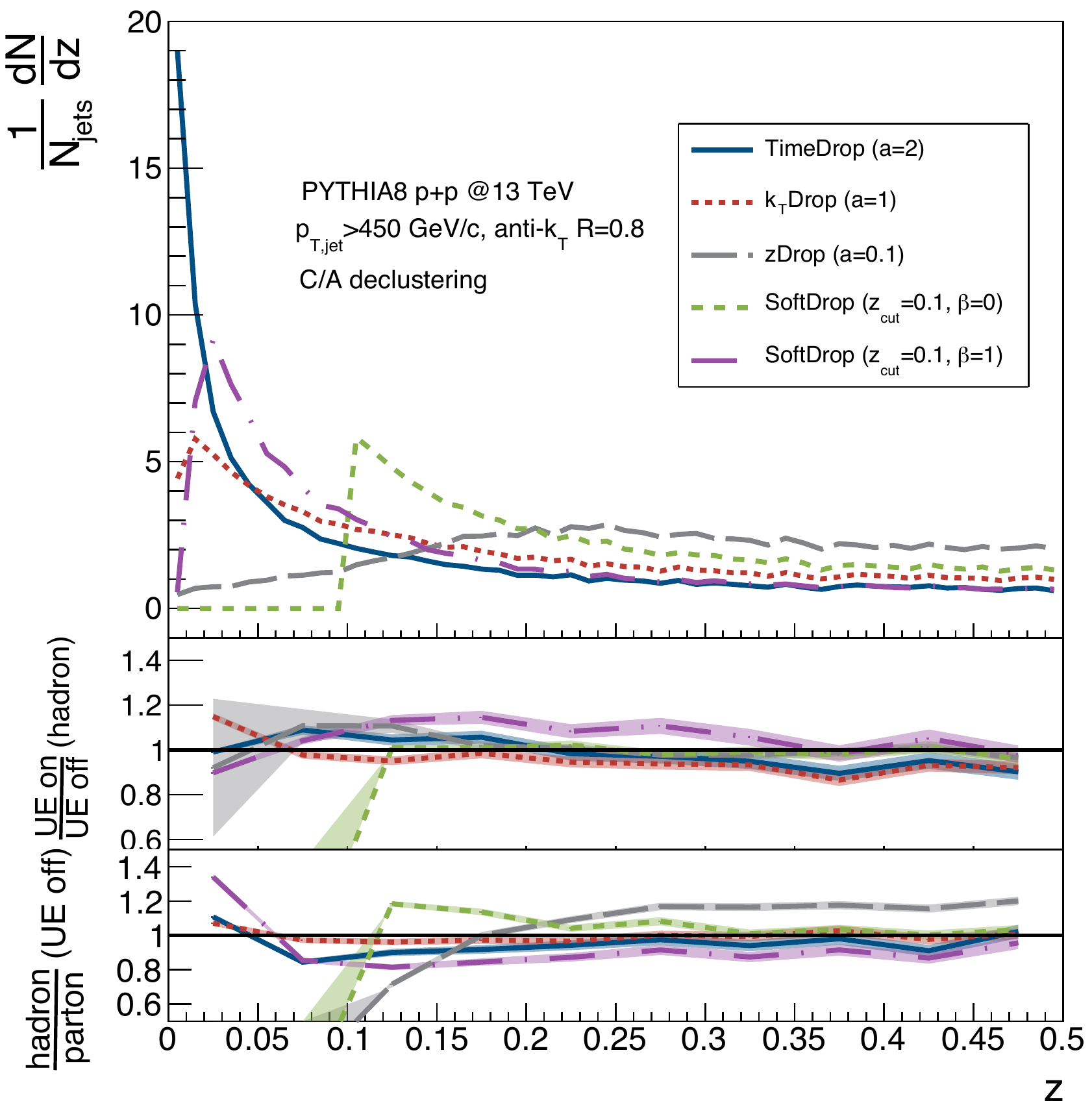}
\caption{Top: the tagged $z$-distribution for different choices of $a$ in Eq.~\ref{eq:hardness} and SoftDrop. Middle: ratio of the distributions with and without underlying event at hadron level. Bottom: ratio of the distributions at hadronic and partonic level without underlying event.}
\label{fig:zdist-mc}
\end{figure}
In this Section, we complement our analytical studies by using PYTHIA8~\cite{Sjostrand:2007gs} to simulate di-jet events in proton-proton collisions at $\sqrt s\!=\!13$~TeV. For each event, particles are clustered into anti-$k_{T}$ jets~\cite{Cacciari:2008gp} with $R\!=\!0.8$ and re-clustered with Cambridge/Aachen using FastJet 3.1~\cite{Cacciari:2011ma}. The analysis is performed on jets with $p_T\!>\!450$~GeV/c. Further, the sensitivity to non-perturbative phenomena such as the underlying event (multi-parton interactions and inital state radiation) and hadronization is explored.

We plot the kinematics of the tagged emissions on the primary Lund plane for the three main choices of $a$ in Eq.~(\ref{eq:hardness}), corresponding to TimeDrop ($a\!=\!2$), $k_t$Drop ($a\!=\!1$) and $z$Drop ($a\!\approx\!0$), in \fign{fig:lundplane-dyg}.  It is clear from these figures that the condition on the hardest branch in each of these three cases corresponds to suppressing the phase space at large formation times (alternatively, small virtualities), small $k_t$'s or small momentum fractions $z$, respectively. It is important to point out that there are no {\it sharp} cuts in the kinematical plane, in contrast to other existing grooming algorithms, such as trimming, filtering, pruning and SoftDrop. This remarkable feature arises due to the fact that the hardest emission, which can be thought of as a proxy of the realistic jet scale, is fluctuating on a jet-by-jet basis. Nevertheless, a dynamical cut is generated which can be estimated by solving $\Delta(\ttau\big|a) \!=\!1/2$. Up to DLA we find
\beq
\kappa= \rme^{-\sqrt{\frac{a \ln 2}{ \abar}}},
\eeq 
or, in terms of $k_t\!=\!z\theta/R$,
\beq
\ln k_{t,\text{cut}}(\theta)= (1-a) \ln \frac{R}{\theta} - \sqrt{\frac{a \ln 2}{ \abar}}. 
\eeq
This defines a straight line boundary that is clear from the MC simulations of the Lund Planes in Fig.~\ref{fig:zdist-mc}. Parametrically,  the point where the critical line crosses the $y$-axis can be estimated within the fixed coupling approximation to scale as
\beq
\ln k_{t,\text{cut}}(R)  \sim - \sqrt{a}. 
\eeq

As an another illustration of the dynamical grooming, we plot in Fig.~\ref{fig:n-hardest} the distribution of the number $i$ of the tagged, hardest branching. Although IRC unsafe it is useful to investigate the location of the tagged branching in the C/A sequence. The larger the power $a$, the more narrow and peaked around $i\!=\!1$, i.e. the first C/A de-clustering step, the distribution is. This is quite natural since $a\!=\!\infty$ corresponds to an angular-ordered Sudakov form factor. In the opposite limit, $a\!\to\!0$, the distribution widens and peaks around $i\!\gg\!1$. More precisely, the average $i$ in each grooming setting is $\approx 2$ (TimeDrop), $3$ ($k_t$Drop) and $5$ ($z$Drop).

The $\rho$-distribution is presented in Fig.~\ref{fig:rhodist-mc}. As anticipated, both TimeDrop and $k_t$Drop exhibit a plain mass-like shape while SoftDrop and $z$Drop deliver an almost flat distribution. The sensitivity to non-perturbative physics is alike for every scenario, especially for values of $\rho\!>\!10^{-3}$. It is worth noticing that while $z$Drop has a fantastic robustness against underlying event, $k_t$Drop outperforms when considering hadronization. Therefore, we expect that a compromise to reduce the sensitivity to both mechanisms simultaneously could be achieved by using intermediate values of $a$ i.e. $1\!>\!a\!>\!0.1$. This possibility together with an extended performance study of the method will be presented in an upcoming publication. 

In the top panel of Fig.~\ref{fig:zdist-mc}, the $z$ distribution of the tagged splitting for different grooming procedures at partonic level and without underlying event is displayed. For completeness, we show the results obtained with SoftDrop for $z_{\rm cut}\!=\!0.1$ and two different choices of $\beta\!=\!0,1$. We find an excellent agreement between the qualitative features of the analytic estimate for the dynamical grooming family as shown in Fig.~\ref{fig:zdist-1} and the more realistic scenario provided by full-fledged Monte-Carlo simulations. It's worth noticing the different behavior between SoftDrop ($\beta\!=\!0$) and $z$Drop even though they use the same variable for tagging, i.e., the momentum sharing. Regarding the impact of non-perturbative effects, in the central and bottom panels we evaluate the role of the underlying event and hadronization, respectively. We would like to highlight the resilience of $k_t$Drop to hadronization effects, an imprint of its effectiveness on selecting the most perturbative splitting. For the other cases, an overall similar performance to SoftDrop is found.  

\section{Conclusions}
\label{sec:conclusions}

In this work we have proposed a new set of jet substructure observables defined by the ``hardest'' splitting in a C/A re-clustered jet. We explore three representative definitions of ''hardness'' in terms of formation time (TimeDrop), relative transverse momentum of the splitting ($k_t$Drop) and momentum sharing ($z$Drop). For a tagged splitting in the shower by either of these three choices, its kinematics serve to compute any observable such as its mass or momentum sharing fraction. Other observables, such as  the groomed angular distribution, can be derived in a completely analogous fashion. 

We have developed an analytical framework that gives a good qualitative understanding of the features seen in full Monte-Carlo simulations. The key object in these calculations is the Sudakov form factor \eqn{eq:sudakov-0} that vetoes all primary emissions in the full angular range of the jet. While many contemporary grooming procedures involve two parameters, our approach relies on the intrinsically generated jet scale whose proxy is the ``hardness'' defined via the continuous parameter $a$. The amount of grooming is somehow related to how different the observable is from the variable that defines the hardness, i.e. by comparing $b$ to $a$, where $a=b$ results in plain distributions. We found also that, in contrast to SD/mMDT where $z_{\rm cut}$ is a parameter of the grooming, a similar cut-off scale is naturally generated by the strong QCD dynamics, $z_{\rm cut} \sim \rme^{-\sqrt{\abar/a}}$. The observables discussed in this work are ICR safe except for the $z$-distribution which  turns out to be Sudakov safe.  

So far we have only investigated observables exploiting the tagged hardest splitting inside a jet. In addition to the remarkable features of the analytic distributions, our Monte-Carlo studies indicate that these observables are quite resilient to non-perturbative effects, including both hadronization and underlying event, for a large part of the distributions. We find it particularly interesting to note that even with relatively mild grooming, $b \lesssim a$, the mass distribution is robust in the region of its peak (this is also the case for the $k_t$ distribution). We propose to study such observables experimentally as they represent, perhaps, the closest realization of perturbative parton dynamics in fully fledged jet observables.

In this work we have deliberately avoided to study in more detail the grooming mode, where branches that violate the ordering set by the hardest branching would be removed leading to modifications of the jet kinematics. This procedure naturally lends itself to an interpretation of removing radiation sensitive to the total color charge of the jet. It  could easily be implemented in a recursive fashion along all the primary and secondary branches/planes of the jet. This will be studied in more detail in an upcoming paper. 

\section*{ACKNOWLEDGMENTS}
We are grateful to Jesse Thaler for a careful reading of the manuscript and useful comments. Y. M.-T. and K. T. acknowledge the ExtreMe Matter Institute, GSI, for hospitality and the participants of the Rapid Reaction Task Force Workshop ``The space-time structure of jet quenching: theory and experiment'' for illuminating discussions.
The work of Y. M.-T. and A. S.-O. was supported by the U.S. Department of Energy, Office of Science, Office of Nuclear Physics, under contract No. DE- SC0012704,
and by Laboratory Directed Research and Development (LDRD) funds from Brookhaven Science Associates. K. T. is supported by a Starting Grant from Trond Mohn Foundation (BFS2018REK01) and the University of Bergen.

\appendix
\section{Collinear unsafe tagging ($a\!=\!0$)}
\label{appendix:a0-case}

The case $a\!=\!0$ has to be treated with special care, since it is formally collinear unsafe. In this case it is also unclear, whether the ``hardest'' emission actually will be one of the primary emissions. Leaving this complication aside for the moment, in this case we would write \eqn{eq:sudakov-mlla-1} as
\begin{align}
\label{eq:sudakov-mlla-0}
\ln \Delta(\ttau\big|0) &= - \int_{0}^1 \dd z' \int_{0}^{R^2} \frac{\dd \theta'^2}{\theta'^2} \, \frac{\alpha_s(z'^2 p^2 \theta'^2)}{2\pi} P(z') \nn
&\times  \Theta\left(z'^2 > z^2\right) \Theta( z'^2 p_{\rm T}^2 \theta'^2 > Q^2_0) \,,
\end{align}
where we have demanded that $k'_t > Q_0$ and $Q_0$ is a non-perturbative scale.
\begin{figure}[h]
\includegraphics[width=\columnwidth]{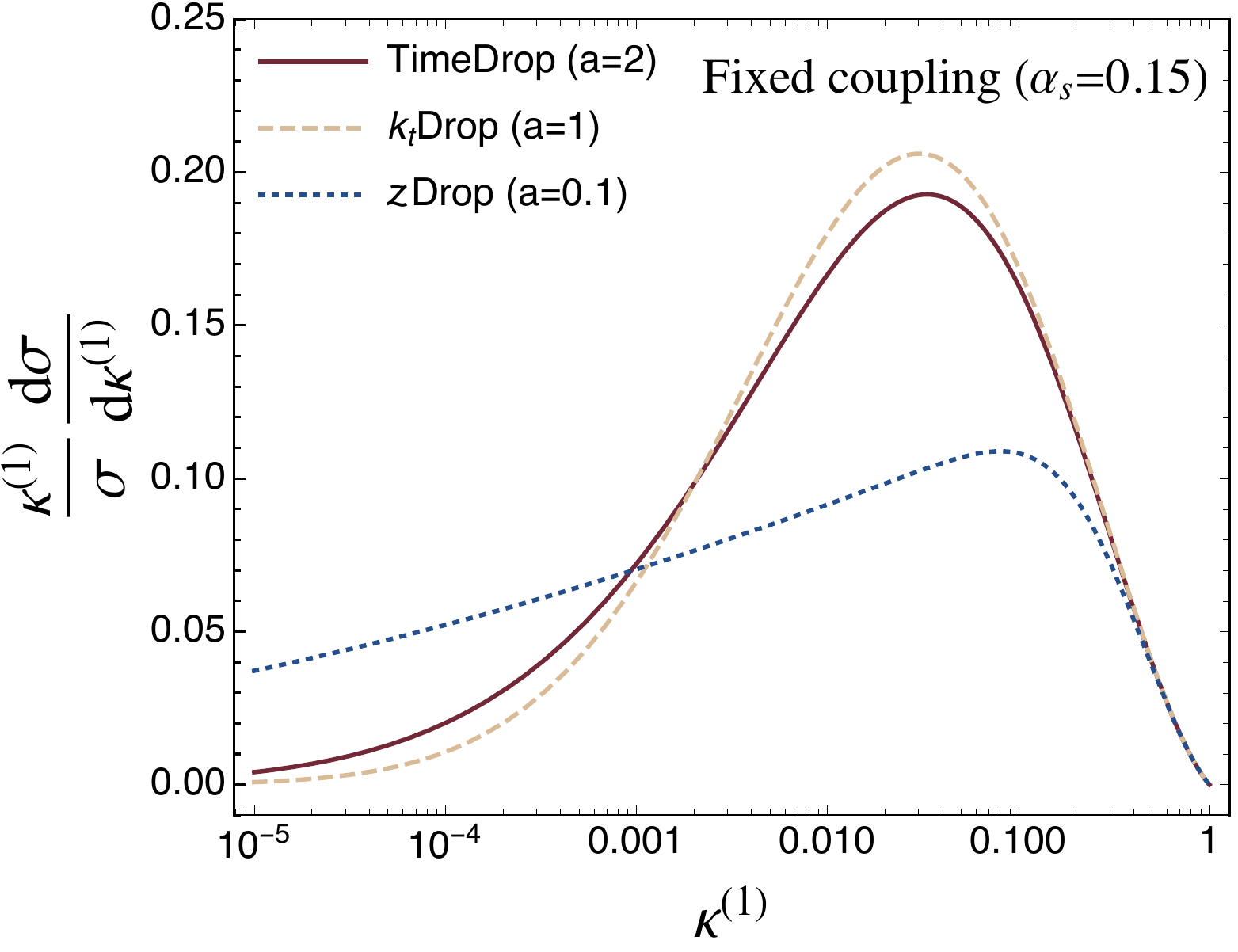}
\caption{The tagged $\ttau^{(1)}$-distribution for fixed coupling as given by \eqn{eq:MLLA-kt} for $2\!>\!a\!>\!0$.}
\label{fig:kt-fc}
\end{figure}

\begin{figure}[ht]
\includegraphics[width=\columnwidth]{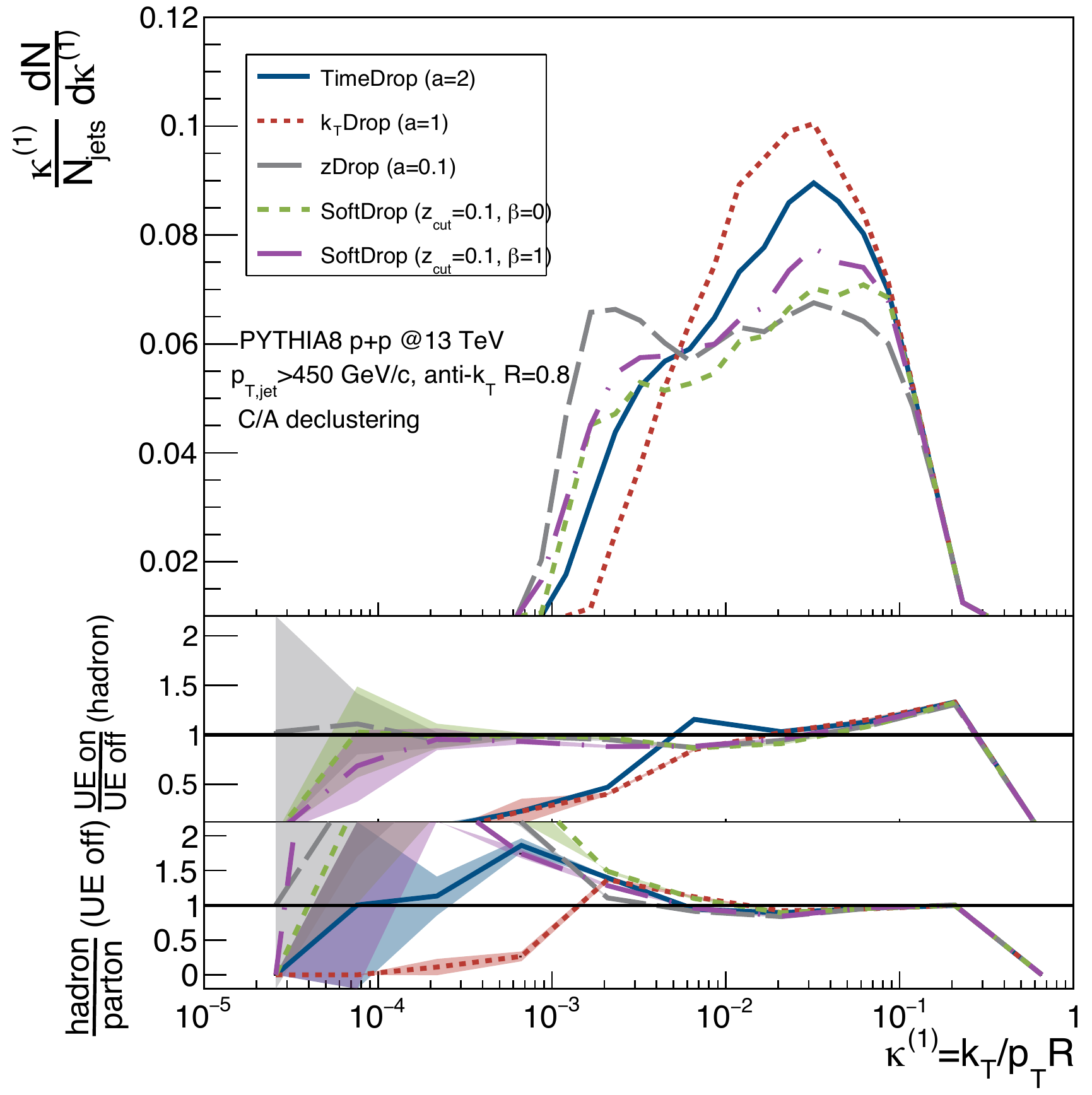}
\caption{Top: the tagged $k_t$-distribution for different choices of $a$ in Eq.~\ref{eq:hardness} and SoftDrop. Middle: ratio of the distributions with and without underlying event at hadron level. Bottom: ratio of the distributions at hadronic and partonic level without underlying event.}
\label{fig:kt-mc}
\end{figure}
For fixed coupling and in DLA, this reads
\begin{align}
\ln \Delta(\ttau\big|0) &= - \bar\alpha \int_z^1 \dd z' P(z') \ln \frac{z' Q}{Q_0} \nn
&\approx -\bar \alpha  \ln\frac{1}{z}\ln\frac{zQ^2}{Q^2_0}.
\end{align}
As a concrete example, let us consider the mass distribution in this case. The expression becomes
\beq
\label{eq:mass-a0}
\frac{m^2}{\sigma} \frac{\dd \sigma}{\dd m^2} = \frac{1}{\cal N} \int_{\max[\frac{m^2}{Q^2},\frac{Q_0^2}{m^2}]}^1 \dd z \, \frac{\alpha_s(z m^2)}{2\pi} P(z) \Delta(\ttau\big|0) \,,
\eeq
where the second condition on the integral comes about by demanding that $k_t^2>Q_0^2$. We notice a strong shape sensitivity to the ratio $Q^2/Q_0^2$. The normalization factor appears from the unitarity condition, and reads
\begin{align}
{\cal N} &= 1- \exp \left[-\int_0^{R^2} \frac{\dd \theta^2}{\theta^2} \int_0^1 \dd z\, \frac{\alpha_s(k^2_t)}{2\pi} P(z) \Theta(k^2_t>Q^2_0) \right] \nn
&\approx 1-\exp \left(-\bar \alpha \ln^2\frac{Q}{Q_0}\right)\,, \nonumber
\end{align}
where the last line was obtained in DLA for fixed coupling.

\section{Tagged $k_t$-distribution ($b\!=\!1$)}
\label{appendix:ktdist}

For completeness, we provide the distribution corresponding to $b\!=\!1$, that is, the relative momentum of the splitting, $\ttau^{(1)} = k_t/Q$. The tagged $\ttau^{(1)}$-distribution is given by
\begin{align}
\label{eq:MLLA-kt}
{\cal H}(1\big|a) & = \frac{2\bar \alpha}{1-a} \int_{\ttau}^{\ttau^{a}} \frac{\dd x}{x}\,\left[1- \frac{3}{4} \frac{x^{\frac{1}{1-a}}}{\ttau^{\frac{a}{1-a}}} \right] \nn
&\times \exp\left\{-\frac{\bar \alpha}{a} \left[\ln^2 x+ \frac{3}{2}(\ln x+1-x) \right] \right\} \,.
\end{align}
After performing the integral in \eqn{eq:MLLA-kt} numerically, we obtain the curves displayed in Fig.~\ref{fig:kt-fc}. In this case, the ordering in the peak position is reverted i.e. it is located at larger values of for $k_t$Drop as compared to TimeDrop. This fact follows naturally from \eqn{eq:hardness-dla-limits}.

Regarding the Monte-Carlo results shown in Fig.~\ref{fig:kt-mc}, a good qualitative agreement with the corresponding analytic formulas is found together with a similar performance among the different scenarios with respect to the impact of the underlying event and hadronization.

\section{Gluon-initiated jet}
\label{appendix:gluon-case}
All along this manuscript we have considered quark-initiated jets as can be deduced from the splitting function given in Eq.~\ref{eq:split-1}. The study of gluon-initiated jets at MLLA amounts to replacing the color factor $C_F$ by $C_A$, and introducing the $g\to g+g$ splitting function
\beq
P(z) = P_{gg}(z) \simeq C_A \left[ \frac{2}{z} - \frac{11}{6} \right] \,,
\eeq
The Sudakov factor then becomes
\begin{align}
\label{eq:sudakov-gluon}
\ln \Delta \big(\ttau \big|a) 
&\simeq -\frac{\abar}{a} \left[\ln^2 \kappa + \frac{11}{6} \left(\ln \kappa + 1 -  \kappa \right) \right] \,,
\end{align}
and the rest of the calculation follows analogously as for the quark.

\begin{figure}[h]
\includegraphics[width=\columnwidth]{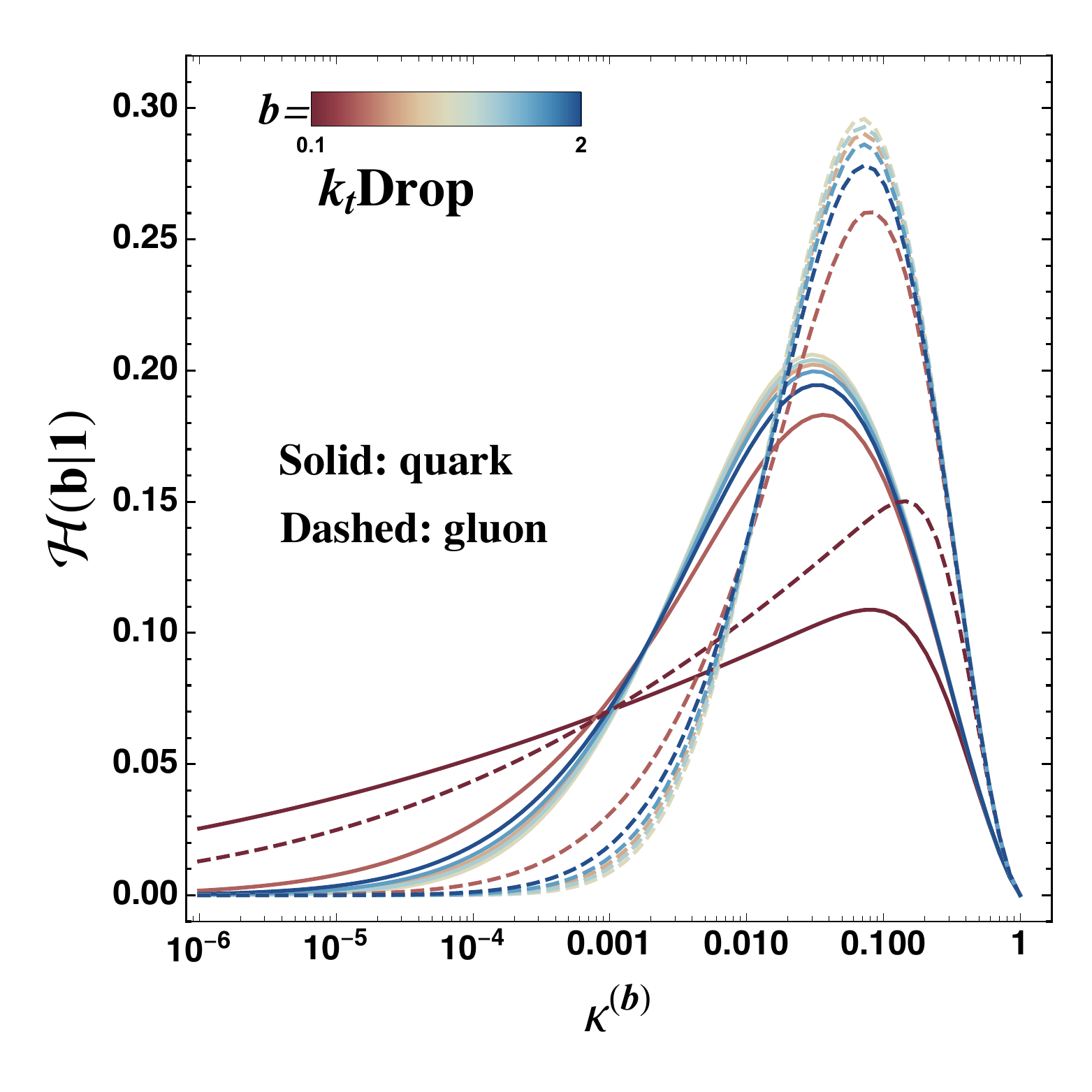}
\caption{The normalized distribution ${\cal H}(b \big| 1)$ for a range of $b$ values in Eq.~\ref{eq:hardness-mlla-0} for fixed coupling at MLLA considering quark (solid) and gluon (dashed) initiated jets. }
\label{fig:gluon_hb1}
\end{figure}

To understand the interplay between the color representation of the initiating parton and the values of $a$ and $b$ in Eq.~\ref{eq:hardness-mlla-0} we have generated an equivalent distribution to the one displayed in the middle panel of Fig.~\ref{fig:Hba-panel} including the gluonic case. The result for ${\cal H}(b \big| 1)$ is shown in Fig.~\ref{fig:gluon_hb1}. The gluon distributions are peaked at larger values of $\kappa$ compared to the quark case as of result of their bigger color charge that translates into an enhanced Sudakov suppression, a well known result~\cite{Larkoski:2013eya}. Note that this transition point occurs at smaller values of $\kappa$ when $b\sim0$.  A more thorough investigation of the quark/gluon-discriminating power of dynamical grooming will be presented in an upcoming publication.

\bibliographystyle{apsrev4-1}
\bibliography{dyg}

\end{document}